\documentclass[preprint,aps]{revtex4}
\usepackage{graphicx}
\usepackage{booktabs}
\usepackage[usenames, dvipsnames]{color}
\usepackage{multirow}
\usepackage{eurosym}
\usepackage{setspace}
\usepackage{rotating}
\usepackage[ pdftex, plainpages = false, pdfpagelabels, 
                 pdfpagelayout = useoutlines,
                 bookmarks,
                 bookmarksopen = true,
                 bookmarksnumbered = true,
                 breaklinks = true,
                 linktocpage=all,
                 pagebackref=false,
                 colorlinks = true,
                 linkcolor = BrickRed,
                 urlcolor  = blue,
                 citecolor = BrickRed,
                 anchorcolor = green,
                 hyperindex = true,
                 hyperfigures
                 ]{hyperref} 
                 
\newcommand{\comment}[1]{}

\begin{document}

\title{Complexity and the Limits of Revolution: \\ What Will Happen to the Arab Spring?}
\author{Alexander S. Gard-Murray and  \href{http://necsi.edu/faculty/bar-yam.html}{Yaneer Bar-Yam}}
\affiliation{\href{http://www.necsi.edu}{New England Complex Systems Institute} \\ 
238 Main St. S319 Cambridge MA 02142, USA}
\date{December 11, 2012}

\begin{abstract}
The recent social unrest across the Middle East and North Africa has deposed dictators who had ruled for decades. While the events have been hailed as an ``Arab Spring'' by those who hope that repressive autocracies will be replaced by democracies, what sort of regimes will eventually emerge from the crisis remains far from certain. Here we provide a complex systems framework, validated by historical precedent, to help answer this question. We describe the dynamics of governmental change as an evolutionary process similar to biological evolution, in which complex organizations gradually arise by replication, variation and competitive selection. Different kinds of governments, however, have differing levels of complexity. Democracies must be more systemically complex
than autocracies because of their need to incorporate large numbers of people in decision-making. This difference has important implications for the relative robustness of democratic and autocratic governments after revolutions. Revolutions may disrupt existing evolved complexity, limiting the potential for building more complex structures quickly. Insofar as systemic
complexity is reduced by revolution, democracy is harder to create in the wake of unrest than autocracy. Applying this analysis to the Middle East and North Africa, we infer that in the absence of stable institutions or external assistance, new governments are in danger of facing increasingly insurmountable challenges and reverting to autocracy.
\end{abstract}

\maketitle

Revolutions can greatly alter societies. Learning about their potential outcomes is important for both participants and policymakers. If we can identify patterns across all revolutions, past unrest may inform our understanding of present crises. Doing so is especially critical as the upheaval across the Middle East and North Africa continues to unfold. 
For a revolution to be successful it must do more than depose or alter the current government; it must also create a new government in line with the intentions of the revolutionaries.  A framework that can provide insight into outcomes is therefore necessary in order to evaluate whether a given revolution is likely to be successful. 

The literature on revolutions has dealt with a wide variety of revolutionary events and outcomes. The conditions cited to explain the outcomes of revolution are themselves very diverse and include, among others, the effect of violence on leader selection \cite{Gurr, BeckerGoldstone}, the dynamics of large organizations \cite{Michels}, the slow nature of state-building \cite{Rueschemeyer}, the rational decisions of participants \cite{WeedeMuller}, elite incentives \cite{AcemogluRobinson}, and leader characteristics \cite{ForanGoodwin}. The definition of revolution ranges widely, from great ``social'' revolutions that reshape society \cite{Skocpol} to lesser ``political'' ones that only change leadership \cite{Goldstone1998}. The diversity of definitions and the range of conditions and factors affecting revolutionary outcomes make it difficult to imagine a general theory that could identify the consequences of present unrest.

As broad and deep as this literature is, it has been criticized by the sociologist Charles Tilly for failing to treat revolutions as ``complex but lawful phenomena'' such as floods or traffic jams. Rather than sui generis events, he argues that they are part of a spectrum of social change, with Òno natural boundariesÓ isolating them from the course of human events. Thus, any definition considering revolutions to be ontologically distinctive is inherently limited.
In this vein, the recent literature is concerned with the specifics of power of groups and institutions that are able to achieve long-range goals \cite{Democratization}.

Here we construct a theory of governmental change from the perspective of complex systems, which can be used to explain and perhaps anticipate the outcomes of revolutions. By embedding social systems in the much broader context of complex systems, we are not limited by the available data on states to formulate hypotheses and frame relevant insights. The available societal data can then serve to validate the mapping of general mathematical principles onto social processes, rather than testing the principles themselves---just as the more commonly applied statistical methods are applicable to physical, biological and social systems and must be used correctly in each case. The focus of our analysis is on the complex challenge of forming a functional government, and the implications of the difficulties in doing so in a revolutionary context.
In this way, complexity theory may provide an additional perspective that may synthesize and extend existing theories about the role that violence, institutions, and time play in revolutions. 
Our theory's applications are not limited to any particular type or scale of revolution. Because it characterizes governmental change generally, it can be applied broadly, with special relevance for periods of rapid and disruptive change such as revolutions. This is helpful when assessing an unfolding event such as the Arab Spring, when it is still unclear whether an event will meet a particular definition of a revolution \cite{Stinchcombe}. In this work we do not consider the causes of revolution themselves \cite{GoldstoneBates,Lagi}. 

The crux of our theory is that revolutions disrupt the complex web of dependencies within governments and between governments and other parts of society, making it more likely that simpler systems, such as autocracy, rather than more complex ones, like democracy (for the current discussion taken to mean representative democracy), will result. We provide empirical support for this theory using data on the outcomes of unrest and governmental changes during the period 1945-2000. In these events higher levels of disruptive violence result in greater incidence of autocratic outcomes. We also show how the sequence of events of widely studied violent revolutions are consistent with our theory, as they revert to autocracy after failing to achieve stable democracy. In evaluating the consistency of objectives of the revolutionaries or observers with likely outcomes, we conclude that whether the Arab Spring will produce democracies hinges on whether existing institutional assets are sufficiently robust to survive the revolutions intact and support the subsequent transition to democracy.

Modern governments are highly complex systems. Complex systems are formed out of many independently acting yet simultaneously interdependent parts. Governments are made up of many people interacting within institutional frameworks (e.g. bureaucrats, judges, and soldiers) equipped with resources such as money, buildings, and weapons, and regulated by laws, norms, procedures, and precedents. It is not merely the quantity of parts, but their interaction and interdependence which give rise to the complexity of the whole.

Highly complex functional systems do not arise spontaneously. In biology, it has long been recognized that complex organisms 
must evolve gradually over many generations from simpler ones; it is probabilistically impossible for random association to produce the necessary relationships between molecules. 
This constraint generally arises in large-scale complex systems such as ecologies, economies, and governments. 
Societal forms do not coincide with biological ones, and the dynamics of social change are not exactly the same as biological evolution. Nevertheless, the constraint that highly complex systems do not arise spontaneously applies to complex social organizations. 
They, too, accumulate structural and functional complexity incrementally 
as described by Darwin's theory of evolution \cite{MTW}. 

The general conditions for an evolutionary process are replication with heredity and variation, and selection with competition \cite{Lewontin}. These conditions can be found in the historical change of governments. Properties of governments are replicated, as new regimes inherit elements of the old and political systems are copied from one country to another; sometimes they are imposed by force, other times they are adopted voluntarily. Innovations and composites of existing governmental forms lead to variation over time.  The innovations are subject to selection through competition between countries. Whether this competition is military, economic, or cultural, national success makes a country's form of government more attractive as a model, as well as better able to impose itself on others via conquest or influence. On the other hand, military or economic failure makes a particular government vulnerable to replacement. This process corresponds to the way genes or traits are selected through the competition between lifeforms:   Less ÒfitÓ governmental forms are more likely to fail, while more fit ones are more likely to succeed and spread. While competitive success may be described in economic and social terms, it is ultimately characterized by the ability to survive and replicate. 

While evolution does not have an inherent objective, the process can result in the progressive development of higher levels of complexity developed from simpler forms \cite{Bonner}. 
We overcome some conceptual difficulties by formalizing complexity in terms of the description of systems at multiple scales  \cite{DCS}. Collective behaviors of components give rise to complex behaviors at larger scales, clarifying the connection between interdependence and complexity.
Thus, new and more complex structures are necessarily dependent on the existence of prior, less complex ones \cite{CXRising}. To the extent that a more complex structure provides a survival advantage, it will be selected for.

Governments, like organisms, may be under evolutionary pressure that results in them becoming more complex. Governmental complexity must increase as the complexity of its environment---the society it governs and other states---increases, in order for Ashby's Law of Requisite Variety to be satisfied \cite{Ashby}. This principle states that in order for a system to survive in a complex environment, its control mechanism must be correspondingly complex. 
The demands and needs of defense, international relations, infrastructure, services, law and order, industrial oversight, and priorities of subpopulations challenge governments.
The large variety of necessary actions must be matched by the ability of the government to act.  
Formally, the complexity of the task as measured by the variety of actions needed to be effective must be matched by the complexity of the organization as measured by the variety of actions that it can make in responding to changing conditions. A complex organization may not be effective, but an effective organization must be sufficiently complex.

A few notes of clarification are necessary. First, the process is not always monotonically increasing in complexity. Complexity can be lost as well as gained, a principle most dramatically demonstrated by historical cases of societal ``collapse''  \cite{Diamond}.  That higher levels of complexity arise based upon previously existing lower levels of complexity does not mean that they always prevail in individual or species competitions. Single celled organisms are necessary for the creation of multicellular organisms, even though bacterial diseases can kill mammals. The fall of Rome might also be considered as a death of a large and more complex organism due to invasion by simpler ones, illustrating how the fate of a single organism, or type, is not ensured by evolutionary process.
Second, while states are sometimes portrayed as being formally planned, the successful implementation of such plans must depend on an incremental accumulation of complexity. Constitutions can be written quickly, but if they call for institutions which are more complex than existing ones, they will take significant time to implement. Third, the process by which complexity increases does not necessarily occur at a steady rate. Complexity may increase more rapidly in one period than previously, as described by the theory of punctuated equilibrium in biological evolution \cite{Goldstone2001,Rueschemeyer}. But even in cases of rapid increase, higher levels of complexity are still dependent on previously existing structures. Fourth, the analysis does not itself place a value on more or less complex organisms or governments, but does provide guidance about the conditions under which certain structures are likely to be successful, and thus the likelihood of achieving the objective if one or another is desired. Fifth, the coexistence of many different types of biological organisms suggests we similarly should not expect a single ultimate or ideal governmental form, and newer forms should continue to emerge. 

The assertion that governmental complexity has increased via incremental evolution is well supported by history. Early political entities, dominated by personalistic hierarchies led by chiefs, kings, and emperors, were far simpler organizations than today's rational-bureaucratic states \cite{Flannery,RichersonBoyd}. Governments gradually increased their complexity by raising armies,  instituting taxes and systems to collect them, building bureaucracies, and so on \cite{Rueschemeyer}. In recent times, the established geography of nations often constrains the evolution of governments in that new options cannot grow organically from smaller successful versions. Instead, variations must happen in situ. This is different from corporations and other social organizations. Both historically and structurally, the complexity of modern states is dependent on the institutions and institutional frameworks of earlier ones. This simplified picture, while necessarily concealing the chaotic and nonlinear nature of history, nevertheless captures the general trend towards building greater complexity on earlier foundations.

Revolutions inherently involve  rapid change that breaks with this cumulative process \cite{ForanGoodwin}. Officials including soldiers, police, judges, and bureaucrats may be removed from office, and both their immediate contributions to government function and their experience may be lost. Laws may be abrogated. Different groups of individuals may become engaged in or empowered by new political arrangements. The more the public sees radical change as necessary, the more likely such disruption of the previous system will be seen as an improvement.

These changes may be considered positive in many regards. The overthrown regime may be corrupt, repressive, or incompetent. The rapid and violent overthrow of even the worst government, however, may still have negative consequences for the development of a new government. By overturning existing structures which could serve as the basis for future structures, revolutions can endanger the very transitions they aim to effect. To the extent that the accumulated complexity of the previous regime is disrupted in the revolution, any new government is forced to build itself without a foundation.

The ongoing needs of society require a functioning government. Moreover, if a new regime is to succeed it must be not merely as effective as the previous one, but given that the old regime's failings prompted a revolution, better. To that end, the post-revolutionary government must restore the lost complexity of essential structures quickly. Revolutions are often prompted by failures of the preexisting regime to meet the basic needs of its people---failures which are often manifest in social crises caused by high food prices,  unemployment, or war. The social  crisis must be addressed promptly in order for any new government to succeed and ward off further unrest. The time available for the government to organize itself and act is therefore short.

These conditions---the loss of complexity and the brief time available to restore it---constrain all post-revolutionary governments. But different forms of government are differentially complex, and less complex ones will fare better under these conditions.

Forms of government may be distinguished in many ways. Among the most important is to whom 
political authority is assigned. Autocracy and democracy represent two very distinct modes of assigning political authority, and the difference between them entails a difference in complexity. This difference in turn affects their robustness when established during a revolution.

An autocratic government is by its nature less complex than a democratic one. Autocracies do not require the extensive decision-making structures that are necessary to gather popular opinion, negotiate with other branches of government, or hold elections or votes in representative bodies. 
Incorporating many independently acting and interacting interests requires special institutions and procedures: legislatures and representatives, competitive elections and debates. 
An autocrat may choose to take these actions 
but they are not required by the form of government.
Many authoritarian governments 
do not conform to the pure autocratic type. Rulers may share power with a group like a family, party, or military \cite{BoixSvolik}. Arrangements like these should entail more political complexity than pure autocracy but need not have the level of complexity necessary for a representative democracy.
Modern democracies also involve a multitude of non-governmental institutions such as civic associations and an independent media. Whatever form these institutions take, the need  to distribute political authority among  many individuals and the diverse groups of which they are a part constitutes an extra layer of complexity within democracy.

In applying our analysis to revolutions that aim to create democratic government we immediately identify the associated difficulties. Having gained power, a provisional government  
must address many questions in order to build its own institutions. There is no single democratic template. The many kinds of democracy---direct and representative, federal and centralized, presidential and parliamentary---reflect its complexity. 
While autocracies may choose embellishments, democracies must determine which of a large number of alternatives to implement. 
Every constitutional choice can become a subject of contention, since each decision might benefit one segment of the population over another.

Quite aside from the problem of structural transformation, the ``complexity gap'' between the institutions of the old regime, minus whatever was lost in the revolution, and those required for the new government to function, are liable to overwhelm attempts to create democracies. 
Even though more complex structures may be capable of responding to higher complexity challenges, the prospective benefits may be stymied by the challenge of creating those structures.

Given these theoretical constraints, we can hypothesize that disruptive revolutionary events will favor the development of autocracies over democracies, even when the impetus of the revolution itself is to create democracy. The degree to which autocracies are favored should increase when there is a greater disruption of the pre-existing governmental form.  The level of disruption may be identified as a first approximation by the level of violence that takes place. The historical record of revolutionary outcomes supports these hypotheses.

In Figure 1 we summarize data from 1945--2000. We track the changes in country regime type in the ten years following revolutionary events. Events are separated into partially overlapping sets based on the particular definition and source used to gather the events. The different sets of events are clustered in three groups. Events that can be characterized as violent revolutions were more likely to produce autocracies. Mass protests and political turmoil were about equally likely to produce democracy as autocracy. More orderly transitions in which leaders voluntarily resigned were the most likely to produce democracy. This is consistent with the hypothesis that violence is more likely to produce autocracy than more peaceful transitions.

While the theory points to the role of disruptive violence itself as a culprit in the instability of democracy after revolution, the data may be consistent with initial conditions playing a causal role in both the extent of violence and the resultant form of governance. Still, the primary conclusion is the same---where violence takes place, the likelihood of autocratic over democratic outcomes is increased. 

\begin{figure}[tb]
	\includegraphics[width=0.8\textwidth]{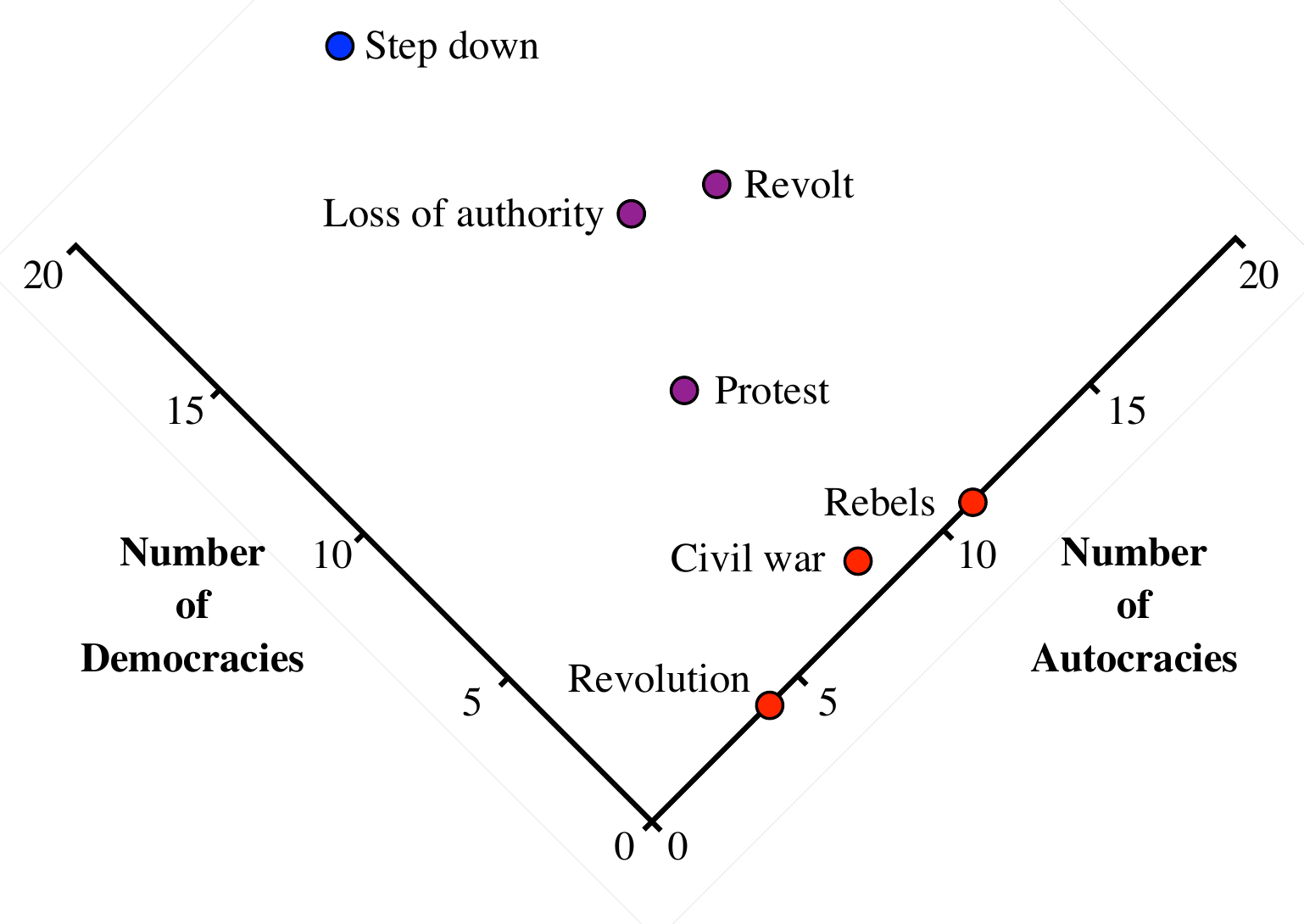}
	\caption{Historical outcomes of governmental changes during the period 1945-2000. Each dot indicates the number of democratic regimes and the number of autocratic regimes from different event collections: Step down, Revolt and Civil war \cite{Svolik};  Protests and Rebels \cite{Archigos}; Loss of authority \cite{Coups}; and Revolution \cite{Polity}. Collections of events with higher levels of organized violence and social disruption are indicated by red dots. Events characterized by mass protest or political turmoil without significant levels of violence are indicated by purple dots. Those in which a leader voluntarily stepped down are indicated by a blue dot. The more violent and disruptive events are more likely to result in autocracies. Regime type is obtained from the Polity project \cite{Polity,Gleditsch}, which characterizes historical governments with daily resolution using multiple coders \cite{MunckVerkuilen}. }
\end{figure}

Though data prior to 1945 is less extensive, we can examine several earlier violent revolutions \cite{Goldstone1998} in greater detail to show how a pattern of revolution, attempted democracy, failed government, and resurgent autocracy repeats itself. The pattern illustrates the difficulty disruptive transitions have in creating stable democratic institutions, even when such institutions are the desired outcome of the revolutionary activity. Through this discussion we also anchor the concepts of the theory in the process of revolution captured in historical narrative, enabling easier application to current events.

The French Revolution of 1789 spawned a series of attempts at representative governments: the National Constituent Assembly, the Legislative Assembly, the National Convention, the Directory. None of these deliberative bodies was able to achieve stability. Instead they were impeded by factional infighting and coups. Dominant groups or leaders drifted towards authoritarianism and repression. Violence and upheaval continued until a military coup replaced the revolutionary governments with autocracy.

It would be easy to blame the failure of the post-revolutionary democratic assemblies and the triumph of autocracy on the ambition of a few autocratically-inclined men like Maximilien Robespierre or Napoleon Bonaparte. But to do so would mask the underlying weakness of the system they subverted. The problem was not that there were no democrats---there was frequent opposition to the autocrats---but that democratic techniques failed to produce governments of sufficient effectiveness to stabilize the country.

The revolution had purged civil institutions of many experienced members, along with the army's aristocratic officer corps \cite{BeckerGoldstone}. The new democratic governments battled civil strife and invaders as well as inflation and hunger, but lacked the apparatus their deposed predecessors possessed. Although elections were held, those elected were never able to hold power for long enough to build the kind of legitimacy that would have discouraged coups. Napoleon, having seized power, could rely on the army as a functioning organization to enable his rule. Whether his policies were good for France can be debated elsewhere; he was nonetheless able to establish an effective and stable government through which to pursue his aims.

Half a century later, Europe was engulfed by the Revolutions of 1848, whose spread mirrored today's Arab Spring in many respects. In France, the monarchy was again overthrown and replaced by a provisional government, which was again unable to resolve the divisions in French politics. Eventually the elected president, Louis-Napoleon Bonaparte (heir to the Emperor), returned the country to autocracy, which brought stability with relatively little popular resistance.

In the Russian Revolution of 1917, the Tsarist autocracy was replaced by a Provisional Government which proclaimed a republic. Democratic processes spread throughout the country, as worker's councils and professional and labor unions formed. But these institutions, while in line with the ideology of the new regime, also diluted its ability to control the state. The Provisional Government suffered from a series of internal crises and cabinet reshuffles, and was unable to prevent the emergence of a competing power center in the form of the Petrograd worker's council. The result was an uneasy system of ``dual power,'' as the Provisional Government's hold on state institutions was gradually weakened in favor of the worker's council (or Soviet). Soon after, the Petrograd Soviet launched a second revolution, starting a civil war that ended four years later with a return to autocracy under the Communist Party.

From these examples a common pattern emerges. First, autocratic regimes are brought down by revolution and new democratic governments are proclaimed. But new legislatures and presidents struggle to meet the challenges before they and their constituents become increasingly dissatisfied. Factionalism emerges as different groups, reacting to and manifesting the weakness of emerging institutions, try to impose their will. Violence erupts, either between the government and the populace or between factions, as immature institutions fail to contain the conflict. Within a few years the new democracies are replaced by regimes generally just as autocratic as the ones recently overthrown, if not more so \cite{Huntington, BeckerGoldstone, WeedeMuller}. 

This process can be translated into the language of complexity: a revolution brings rapid change, displacing or destroying accumulated complexity. Subsequently, governments try to create democracy, but find the gap between their present and intended complexity is too great and are replaced by autocracy. The problem is not that democracy is flawed; the stability and prosperity of mature democratic states testifies to that. But the greater complexity of democracy means that it is harder to build when time is short and the foundations for more complex structures have been eroded or did not exist in the first place.

Not all revolutions end in autocracy. The extent of disruption plays a pivotal role. Particularly in the last several decades, several revolutions---in the Philippines in 1986, in South Korea in 1987, and across Eastern Europe in 1989---have managed to oust autocratic regimes and replace them with more or less stable democracies. These are not counter-examples, but rather demonstrations of the constructive side of our conclusions. In all of these nations, the revolutions were largely non-violent and forced out autocrats without overturning the apparatus of government.  These revolutions also managed to take advantage of the existing store of complexity represented by previously powerless democratic elements, such as parliaments with only symbolic roles and tightly managed elections.  Put in place by autocratic governments to provide the semblance of representational government, these powerless institutions later provided a basis for a democratic government without significant changes to their structure.
Furthermore, these regimes all received external support  
during their critical early years. The retention of complexity in the form of previous government structures, combined with external support, allowed democracy to grow incrementally and stabilize.

Even in violent revolutions, maintaining existing complexity can help support post-revolutionary democracy. 
In the American Revolution, the groundwork for democracy was laid prior to 1776 in elected colonial assemblies. When the Articles of Confederation outlined a federal government for the new nation, they were able to build on functioning democratic foundations that had existed for decades. The responsibilities of the federal government could then grow over time, while the state governments continued to address the needs of their populations as they had done previously.

For today's revolutionaries in the Middle East, the complexity of governmental formation does not mean democracy is impossible, but it does mean that it will be very hard to create. Our analysis does not incorporate distinct cultural, historical, economic and other conditions, but the fundamental principles should provide helpful insight. With regards to Egypt and Tunisia, many observers are concerned that the pace of democratization has been slow and the old regime remains essentially in place. But this path, while it risks the persistence of the old guard, may offer the best hope for a smooth transition. Preserving and gradually changing some of the existing structures of government may ultimately be more likely to succeed. Efforts to dramatically accelerate change may cause a reversion to the prior form of autocratic government, or extended social disorder. 

Contrast this with Libya and Syria, where the revolutions have been far more violent, threatening the prospects for a smooth transition. Libya already faces emerging fragmentation within the National Transition Council. 
The conflict created many militias which may be competitors for power in the absence of a strong and accepted government.
The same weapons which overthrew Muammar Qaddafi may now fuel violent conflict between revolutionary and ethnic factions. In Syria, severe repression means much of the old regime's army and bureaucracy will be too discredited to participate in a new government. Bashir Assad's determination to maintain power regardless of the level of civilian casualties, and the apparent loyalty of his security apparatus,  means that if the revolution does succeed it will have to destroy much of the Assad regime's institutions, leaving little foundation on which to build a complex, democratic post-Assad government.

The preceding analysis has focused on the internal dynamics of new governments. External interventions, whether through aid or military action, have the potential to alter these dynamics. In Libya, NATO airstrikes contributed to the rebels' victory, though how external interventions might promote the formation of a stable government is less straightforward. Saudi Arabian military intervention in Bahrain stopped the incipient revolution there. Still, the potential of foreign involvement should not be overstated. The relevance of diverse local imperatives in shaping the creation of complex democratic institutions cautions against the external imposition of such institutions. 

The likelihood of a persistent democracy can be increased if new internal disruptions are inhibited by external intervention. First, the political structure must be stabilized and economic stresses reduced, allowing governments time to develop. All such external interventions must, however, be carefully considered.  Evolution is always shaped by the environment: national institutions that develop in the presence of external aid may become reliant on it.  Thus, the form of assistance should be carefully designed to be like scaffolding---to be removed rather than to become integral to the functioning of the system \cite{MTW}.

A separate category of constructive intervention is the prevention of global conditions that increase stress on vulnerable countries. In this regard, high and volatile global food prices are a key ongoing culprit for social unrest and political instability that can and should be addressed \cite{Lagi}. 

Aside from such interventions, the fundamental dynamics of building complex systems like governments constrain the outcomes of revolutions. Even if external interventions do occur their success is far from assured. Governments and societies are evolving systems which develop over long periods of time. Stability can neither be assumed nor instantly restored after revolution. Recognizing the difficulties that revolutions create for post-revolutionary governments and societies may help guide our response to social unrest. Building a successful government begins, rather than ends, with the revolution. 

\section*{Acknowledgments}
\label {sec:ackn}

We thank Greg Lindsay, Karla Z. Bertrand, Dominic Albino and Yavni Bar-Yam for editorial assistance, Lawrence E Susskind, Robert H. Bates and Dietrich Rueschemeyer for helpful comments on the manuscript. This work was supported in part by AFOSR under grant FA9550-09-1-0324, ONR under grant N000140910516.


\begin{thebibliography}{10}

\bibitem{Gurr}
T.~R. Gurr, War, revolution, and the growth of the coercive state, {\it
  Comparative Political Studies\/} {\bf 21}, 45 (1988).

\bibitem{BeckerGoldstone}
J.~Becker, J.~A. Goldstone, {\it \emph{State development after revolutions -
  Rapid state building or transforming existing structures under pressure?}\/},
  in~\emph{States and Development: Historical Antecedents of Stagnation and
  Advance} M{.}~Lange, D.~Rueschemeyer, eds. (Palgrave Macmillan, 2005), pp.
  183--210.

\bibitem{Michels}
R.~Michels, {\it Political Parties: A Sociological Study of the Oligarchical
  Tendencies of Modern Democracy\/} (Batoche Books, 2001).

\bibitem{Rueschemeyer}
D.~Rueschemeyer, {\it \emph{Building states: Inherently a long-term process? An
  argument from theory}\/}, in~\emph{States and Development: Historical
  Antecedents of Stagnation and Advance} M{.}~Lange, D.~Rueschemeyer, eds.
  (Palgrave Macmillan, 2005), pp. 143--164.

\bibitem{WeedeMuller}
E.~Weede, E.~N. Muller, Consequences of revolutions, {\it Rationality and
  Society\/} {\bf 9}, 327 (1997).

\bibitem{AcemogluRobinson}
D.~Acemoglu, J.~A. Robinson, {\it Economic origins of dictatorship and
  democracy\/} (Cambridge University Press, 2006).

\bibitem{ForanGoodwin}
J.~Foran, J.~Goodwin, Revolutionary outcomes in Iran and Nicaragua: Coalition
  fragmentation, war, and the limits of social transformation, {\it Theory and
  Society\/} {\bf 22}, 209 (1993).

\bibitem{Skocpol}
T.~Skocpol, {\it States and Social Revolutions: A Comparative Analysis of
  France, Russia, and China\/} (Cambridge University Press, 1979).

\bibitem{Goldstone1998}
J.~A. Goldstone, {\it The Encyclopedia of Political Revolutions\/}
  (Congressional Quarterly, 1998).

\bibitem{Democratization}
D.~Rueschemeyer, On the state and prospects of comparative democratization
  research, {\it Comparative Democratization Section of the American Political
  Science Association (APSA-CD) Newsletter\/} {\bf 8}, 1 (October, 2010).

\bibitem{Stinchcombe}
A.~L. Stinchcombe, Ending revolutions and building new governments, {\it Annual
  Review of Political Science\/} {\bf 2}, 49 (1999).

\bibitem{GoldstoneBates}
J.~A. Goldstone, R.~H. Bates, D.~L. Epstein, T.~R. Gurr, M.~B. Lustik, M.~G.
  Marshall, J.~Ulfelder, M.~Woodward, A global model for forecasting political
  stability, {\it American Journal of Political Science\/} {\bf 54}, 190
  (2010).

\bibitem{Lagi}
M.~Lagi, K.~Z. Bertrand, Y.~Bar-Yam, The food crises and political instability
  in North Africa and the Middle East  (2011).

\bibitem{MTW}
Y.~Bar-Yam, {\it Making Things Work: Solving Complex Problems in a Complex
  World\/} (NECSI Knowledge Press, 2005).

\bibitem{Lewontin}
R.~C. Lewontin, The units of selection, {\it Annual Review of Ecology and
  Systematics\/} {\bf 1}, 1 (1970).

\bibitem{Bonner}
J.~T. Bonner, {\it The Evolution of Complexity by Means of Natural Selection\/}
  (Princeton University Press, 1988).

\bibitem{DCS}
Y.~Bar-Yam, {\it Dynamics of Complex Systems\/} (Westview Press, 1997).

\bibitem{CXRising}
Y.~Bar-Yam, {\it Complexity rising: From human beings to human civilization, a
  complexity profile, \emph{Encyclopedia of Life Support Systems}\/}
  (UNESCO/EOLSS Publishers, 2002).

\bibitem{Ashby}
W.~R. Ashby, {\it An Introduction to Cybernetics\/} (Chapman \& Hall, 1956),
  chap.~11.

\bibitem{Diamond}
J.~Diamond, {\it Collapse: How Societies Choose to Fail or Succeed\/} (Viking
  Penguin, 2005).

\bibitem{Goldstone2001}
J.~A. Goldstone, Toward a Fourth Generation of Revolutionary Theory, {\it
  Annual Review of Political Science\/} {\bf 4}, 139 (2001).

\bibitem{Flannery}
K.~V. Flannery, The cultural evolution of civilizations, {\it Annual Review of
  Ecology and Systematics\/} {\bf 3}, 399 (1972).

\bibitem{RichersonBoyd}
P.~J. Richerson, R.~Boyd, Complex societies: The evolutionary origins of a
  crude superorganism, {\it Human Nature\/} {\bf 10}, 253 (1972).

\bibitem{BoixSvolik}
C.~Boix, M.~Svolik, The foundations of limited authoritarian government:
  Institutions and power-sharing in dictatorships, {W}orking paper (2011).

\bibitem{Svolik}
M.~Svolik, Power-sharing and leadership dynamics in authoritarian regimes, {\it
  American Journal of Political Science\/} {\bf 53}, 477 (2009).

\bibitem{Archigos}
H.~E. Goemans, K.~S. Gleditsch, G.~Chiozza, Introducing Archigos: A data set of
  political leaders, {\it Journal of Peace Research\/} {\bf 46}, 269 (2009).

\bibitem{Coups}
M.~G. Marshall, D.~R. Marshall, Coup d'Etat Events, 1946-2010: Codebook, {\it
  Center for Systemic Peace\/}  (2011).

\bibitem{Polity}
M.~G. Marshall, K.~Jaggers, T.~R. Gurr, Polity {IV} Project: {P}olitical regime
  characteristics and transitions, 1800-2010, {\it Center for Systemic Peace\/}
   (2012).

\bibitem{Gleditsch}
K.~S. Gleditsch, Modified polity {P4} and {P4D} data, version 3.0,
  \url{http://privatewww.essex.ac.uk/~ksg/polity.html} (2008).

\bibitem{MunckVerkuilen}
G.~L. Munck, J.~Verkuilen, Conceptualizing and measuring democracy: Evaluating
  alternative indices, {\it Comparative Political Studies\/} {\bf 35}, 5
  (2002).

\bibitem{Huntington}
S.~P. Huntington, {\it Political Order in Changing Societies\/} (Yale
  University Press, 1968).

\end{thebibliography}
\end{document}